\documentclass[pdflatex,sn-aps]{sn-jnl}

\usepackage{graphicx}
\usepackage{multirow}
\usepackage{amsmath,amssymb,amsfonts}
\usepackage{amsthm}
\usepackage{mathrsfs}
\usepackage[title]{appendix}
\usepackage{xcolor}
\usepackage{textcomp}
\usepackage{manyfoot}
\usepackage{booktabs}
\usepackage{algorithm}
\usepackage{algorithmicx}
\usepackage{algpseudocode}
\usepackage{listings}

\usepackage{hyperref}
\usepackage{cleveref}
\usepackage{xspace}
\usepackage{xstring}
\usepackage{setspace}

\allowdisplaybreaks

\DeclareRobustCommand{\ccite}[1]{\IfSubStr{#1}{,}{refs.~}{ref.~}\cite{#1}}
\DeclareRobustCommand{\Ccite}[1]{\IfSubStr{#1}{,}{Refs.~}{Ref.~}\cite{#1}}

\newcommand{\eg}{\textit{e.g.}\xspace}
\newcommand{\ie}{\textit{i.e.}\xspace}

\newcommand{\sn}[5]{\ensuremath{#1
    \scalebox{0.7}{\ensuremath{\left[^{#2}{\kern -0.15em}{#3}_{#4}
          \ifthenelse{\equal{#5}{}}{}{{}^{(#5)}}\right]}}}\xspace}
\newcommand{\snq}[5]{\sn{\Pqqbar[#1]}{#2}{#3}{#4}{#5}}
\newcommand{\sno}[5]{\ensuremath{\langle\sn{\mathcal{O}_{#1}}{#2}{#3}{#4}{#5}
    \rangle}\xspace}

\newcommand{\dif}[1][]{\ensuremath{\ifthenelse{\equal{#1}{}}
    {\,\text{d}}{\text{d}#1\,}}}

\newcommand{\ket}[1]{\left|#1\right\rangle}
\newcommand{\mmp}[1]{\,#1}
\renewcommand{\max}{\ensuremath{\text{max}}\xspace}
\renewcommand{\min}{\ensuremath{\text{min}}\xspace}

\newcommand{\pss}[1][]{\ensuremath{\Delta_{#1}}\xspace}
\newcommand{\psk}[1][]{\ensuremath{P_{#1}}\xspace}
\newcommand{\pso}[1][]{\ensuremath{O_{#1}}\xspace}
\newcommand{\psd}[1]{\ensuremath{D_{#1}}\xspace}
\newcommand{\pmu}{\ensuremath{\mu}\xspace}

\newcommand{\pythia}[1][8]{\textsc{Pythia}~#1\xspace}
\newcommand{\ariadne}{\textsc{Ariadne}\xspace}

\newcommand{\Ppsi}[1][1]{\ensuremath{\ifthenelse{\equal{#1}
      {}}{{\psi}}{\psi(#1S)}}\xspace}
\newcommand{\Peta}[1][]{\ensuremath{\ifthenelse{\equal{#1}
      {}}{\eta_c}{\eta_{#1}}}\xspace}
\newcommand{\Pchi}[1][]{\ensuremath{\ifthenelse{\equal{#1}
      {}}{\chi_{c}}{\chi_{#1}}}\xspace}
\newcommand{\Pups}[1][]{\ensuremath{\ifthenelse{\equal{#1}
      {}}{\Upsilon}{\Upsilon(#1S)}}\xspace}
\newcommand{\Pq}[1][q]{\ensuremath{#1}\xspace}
\newcommand{\Pqbar}[1][q]{\ensuremath{\overline{\Pq[#1]}}\xspace}
\newcommand{\Pqqbar}[1][q]{\ensuremath{\Pq[#1]\Pqbar[#1]}\xspace}
\newcommand{\Pc}{\Pq[c]}

\newcommand{\Pccbar}{\Pqqbar[c]}
\newcommand{\Pb}{\Pq[b]}

\newcommand{\Pbbbar}{\Pqqbar[b]}
\newcommand{\PQ}{\Pq[Q]}

\newcommand{\PQQbar}{\Pqqbar[Q]}
\newcommand{\Pg}{\ensuremath{g}\xspace}

\newcommand{\PZ}{\ensuremath{Z}\xspace}

\newcommand{\gev}{\ensuremath{\mathrm{\,Ge\kern -0.1em V}}\xspace}
\newcommand{\tev}{\ensuremath{\mathrm{\,Te\kern -0.1em V}}\xspace}

\newcommand{\as}{\ensuremath{\alpha_s}\xspace}
\newcommand{\pt}{{\ensuremath{p_\text{T}}}\xspace}
\newcommand{\pte}[1][]{{\ensuremath{p_\text{T,evol}^{#1}}}\xspace}
\newcommand{\com}{\ensuremath{\sqrt{s}}\xspace}

\newcommand{\reportnumber}{FERMILAB-PUB-23-761-CSAID}
\usepackage[angle=0,scale=1,color=black,firstpage=true,opacity=1]{background}
\SetBgContents{\footnotesize\sf{\reportnumber}}
\SetBgPosition{current page.north east}
\SetBgHshift{-2in}
\SetBgVshift{-0.5in}

\begin{document}

%%%%%%%%%%%%%%%%%%%%%%%%%%%%%%%%%%%%%%%%%%%%%%%%%%%%%%%%%%%%%%%%%%%%%%%%%%%%%%%
\title{Non-Relativistic Quantum Chromodynamics in Parton Showers}

% Authors.
\author[1]{\fnm{Naomi} \sur{Cooke}}
\author[2]{\fnm{Philip} \sur{Ilten}}
\author[3]{\fnm{Leif} \sur{L\"onnblad}}
\author[2,4]{\fnm{Stephen} \sur{Mrenna}}
\affil[1]{
  \orgdiv{School of Physics and Astronomy},
  \orgname{University of Glasgow},
  \orgaddress{\city{Glasgow}, \postcode{G12 8QQ}, \country{United Kingdom}}}
\affil[2]{
  \orgdiv{Department of Physics},
  \orgname{University of Cincinnati},
  \orgaddress{\city{Cincinnati}, \state{OH} \postcode{45221}, \country{USA}}}
\affil[3]{
  \orgdiv{Department of Physics},
  \orgname{Lund University},
  \orgaddress{\postcode{SE-221 00} \city{Lund}, \country{Sweden}}}
\affil[4]{
  \orgdiv{}
  \orgname{Fermi National Accelerator Laboratory},
  \orgaddress{\city{Batavia}, \state{IL} \postcode{60510}, \country{USA}}}

% Abstract.
\abstract{Measurements of quarkonia isolation in jets at the Large Hadron Collider (LHC) have been shown to disagree with fixed-order non-relativistic quantum chromodynamics (NRQCD) calculations, even at higher orders. Calculations using the fragmenting jet function formalism are able to better describe data but cannot provide full event-level predictions. In this work we provide an alternative model via NRQCD production of quarkonia in a timelike parton shower. We include this model in the \pythia event generator and validate our parton-shower implementation against analytic forms of the relevant fragmentation functions. Finally, we make inclusive predictions of quarkonia production for the decay of the standard-model Higgs boson.}
\maketitle

%%%%%%%%%%%%%%%%%%%%%%%%%%%%%%%%%%%%%%%%%%%%%%%%%%%%%%%%%%%%%%%%%%%%%%%%%%%%%%%
\section{Introduction}\label{sec:intro}

The production of quarkonia, flavourless mesons comprised of a heavy quark and its corresponding antiquark bound by the strong interaction, is still not fully understood in the context of hadron-hadron collisions, despite decades of study~\cite{Clark:1978mg, Ueno:1978vr,  UA1:1987azc, UA1:1988pyx, UA1:1990eni, CDF:1997uzj, CDF:2004jtw, D0:1996awi, LHCb:2015foc, LHCb:2013itw, LHCb:2012kaz, LHCb:2011zfl, ATLAS:2015zdw, CMS:2015lbl, CDF:2007msx, LHCb:2013izl, ALICE:2011gej, CMS:2013gbz, LHCb:2017llq, CMS:2019ebt, CMS:2021puf}. For a complete review, see \ccite{Lansberg:2006dh}. The reason for this is that our theory of the strong interaction, quantum chromodynamics (QCD), is best understood at large energy scales, whereas the dynamics of quarkonia production spans both large and small scales.

One of the first models of quarkonium production to make quantitative predictions was the colour-singlet model (CSM)~\cite{Chang:1979nn, Baier:1981uk, Baier:1983va, Halzen:1984rq, Glover:1987az}, which factorises these perturbative and non-perturbative regimes. In the CSM, the production of the heavy quark/antiquark pair (\PQQbar) is calculated using perturbative QCD, and these quarks are then bound non-perturbatively. In this second step, the constituents are assumed to be at rest in the quarkonium frame, and the colour and spin state of the \PQQbar pair does not change during the binding. Comparisons of CSM predictions to Tevatron data~\cite{CDF:2004jtw, D0:1996awi} demonstrated that the CSM, when calculated at leading fixed order, significantly underestimated the prompt production of \Ppsi mesons, a \snq{\Pc}{3}{S}{1}{} quarkonium state\footnote{Here, spectroscopic notation is being used of the form \snq{\PQ}{2S+1}{L}{J}{}, where \PQ is the heavy flavour quark/antiquark pair, and $S$, $L$, $J$ are the total spin, orbital, and total angular-momentum quantum numbers, respectively. For $L$, the orbitals $S$, $P$, and $D$ correspond to orbital angular-momentum quantum numbers of $0$, $1$, and $2$, respectively.}.

The initial fixed-order CSM predictions~\cite{Glover:1987az} for the Tevatron data lacked two important sources of \Ppsi production: one, partonic fragmentation; and, two, feed-down from heavier quarkonium states such as the \sn{}{3}{P}{J}{} \Pchi mesons. This first contribution can be significant in hadronic collisions because the rate for inclusive charm production is much larger than that calculated only from fixed-order processes such as $\Pqqbar \to \Pccbar$ and $\Pg\Pg \to \Pccbar$. Fragmentation functions~\cite{Braaten:1994xb, Braaten:1993rw, Braaten:1993mp, Braaten:1994kd, Cacciari:1994dr} were derived for $\Pc \to \snq{\Pc}{3}{S}{1}{} \Pc$, considering the simpler $\PZ \to \Pccbar$ process, and convolved with Tevatron jet cross-sections. This additional fragmentation contribution brought the CSM prediction into better agreement with data at high \pt of the \Ppsi, but the overall agreement with data was still poor. For feed-down contributions from the \Pchi, predictions were made at fixed-order with the CSM~\cite{Barbieri:1979iy, Barbieri:1980yp, Barbieri:1981xz}. These calculations contain a kinematic cut-off to regulate an infrared divergence, typically taken as the mass of the bound state. The quantitative CSM results are very sensitive to the exact choice of this cut-off. Even after accounting for feed-down, the CSM prediction still significantly disagrees with the Tevatron data.

With the realization that the CSM was inadequate to describe the data, non-relativistic QCD (NRQCD) theory~\cite{Bodwin:1994jh, Braaten:1994vv, Cho:1995vh, Cho:1995ce} was developed. While NRQCD is similar to the CSM in that it factorises into a perturbative and non-perturbative component, it does not require the initial partonic state to have the same colour or quantum numbers as the final state, and allows, for example, the production of colour-octet states which then evolve into physical colour-singlet states.

Production rates for quarkonia at both the Tevatron and LHC are, in general, well described by NRQCD. However, NRQCD predicts \Ppsi polarisation~\cite{Leibovich:1996pa, Braaten:2000gw, Beneke:1996yw, Braaten:1999qk, Gong:2012ug, Chao:2012iv} that is not observed in LHC data~\cite{LHCb:2013izl, ALICE:2011gej, CMS:2013gbz}. Additionally, fixed-order NRQCD predicts relatively isolated \Ppsi production, while LHC measurements demonstrate that \Ppsi mesons are oftentimes produced in association with nearby particles. Recent developments in calculating quarkonium fragmentation functions promise to resolve these issues~\cite{Bain:2016rrv, Bain:2017wvk}. However, a comprehensive approach for calculating these fragmentation functions is lacking. Here, we introduce an alternative approach of using the \pythia~\cite{Sjostrand:2006za, Sjostrand:2007gs, Sjostrand:2014zea, Bierlich:2022pfr} parton shower to calculate the production of quarkonia from fragmentation. The idea is inspired by an earlier study using the \ariadne program
\cite{Lonnblad:1992tz, Ernstrom:1996am, Ernstrom:1996aa}, but is implemented here with a complete and extensible set of quarkonia processes in the full \pythia framework.

The remainder of this paper is organised as follows. In \cref{sec:nrqcd} we briefly review NRQCD, while in \cref{sec:shower} we describe parton shower formalism. We then include NRQCD quarkonium fragmentation processes into the \pythia parton-shower in \cref{sec:imp}. In \cref{sec:results}, we present quantitative predictions using our approach, and finally, we provide conclusions in \cref{sec:con}.

%%%%%%%%%%%%%%%%%%%%%%%%%%%%%%%%%%%%%%%%%%%%%%%%%%%%%%%%%%%%%%%%%%%%%%%%%%%%%%%
\section{Non-relativistic Quantum Chromodynamics}\label{sec:nrqcd}

Non-relativistic QCD is an effective field theory where a perturbative expansion is performed not only in the QCD coupling \as, but also in the relative velocity between the two heavy quarks of the bound state. Including this relative velocity scale results in the expansion of the physical quarkonium state into non-physical Fock states. For a heavy quark state \PQQbar, the general Fock-state expansion is,
\begin{equation}
  \begin{aligned}
    \ket{\sn{\PQQbar}{2S+1}{L}{J}{}} ={}
    & \mathcal{O}(1)\ket{\snq{\PQ}{2S+1}{L}{J}{1}} +
    \mathcal{O}(v)\ket{\snq{\PQ}{2S+1}{L\pm1}{J'}{8}\Pg} \\
    & + \mathcal{O}(v^2)\ket{\snq{\PQ}{2S+1}{L}{J}{8}\Pg\Pg} + \ldots \mmp{,}
    \label{equ:fock}
  \end{aligned}
\end{equation}
where $v$ is the relative velocity between the constituent quark/antiquark, $(1)$ in the superscript notates a colour-singlet state, and $(8)$ in the superscript notates a colour-octet state. For a \Ppsi, this can be explicitly written as,
\begin{equation}
  \begin{aligned}
    \ket{\sn{\Pccbar}{3}{S}{1}{}} ={}
    & \mathcal{O}(1)\ket{\snq{\Pc}{3}{S}{1}{1}}
    + \mathcal{O}(v)\ket{\snq{\Pc}{3}{P}{J}{8}\Pg}
    + \mathcal{O}(v^2)\ket{\snq{\Pc}{1}{S}{0}{8}\Pg} \\
    & + \mathcal{O}(v^2)\ket{\snq{\Pc}{3}{S}{1}{8}\Pg\Pg} + \ldots \mmp{,}
  \end{aligned}
  \label{equ:fockPsi}
\end{equation}
and for the \Pchi states as,
\begin{equation}
  \ket{\sn{\Pccbar}{3}{P}{J}{}} =
  \mathcal{O}(1)\ket{\snq{\Pc}{3}{P}{J}{1}}
  + \mathcal{O}(v)\ket{\snq{\Pc}{3}{S}{1}{8}g} + \ldots \mmp{,}
  \label{equ:fockChi}
\end{equation}
where $J$ corresponds to $0$, $1$, and $2$.

The heavy quarkonium cross-sections can then be factorised into short- and long-distance scales, which for initial partons $a_1$ and $a_2$ can be written as
\begin{equation}
  \begin{aligned}
    & \dif\hat{\sigma}(a_1 a_2 \to \snq{\PQ}{2S+1}{L}{J}{} X) =
    \sum_{S'} \sum_{L'} \sum_{J'} \sum_{C'} \\
    & \quad \sno{\snq{\PQ}{2S+1}{L}{J}{}}{2S'+1}{L'}{J'}{C'}
    \dif\hat{\sigma}(a_1 a_2 \to \snq{\PQ}{2S'+1}{L'}{J'}{C'} X) \mmp{,}
  \end{aligned}
  \label{equ:sigmaHat}
\end{equation}
where the $\dif\hat{\sigma}(a_1 a_2 \to \snq{\PQ}{2S'+1}{L'}{J'}{C'} X)$ are perturbatively calculated partonic short-distance matrix elements (SDME) and \sno{\snq{\PQ}{2S+1}{L}{J}{}}{2S'+1}{L'}{J'}{C'} are empirically determined long-distance matrix elements (LDME). The sums are over the states of the Fock expansion of \cref{equ:fock}, where the indices correspond to the spin and colour quantum numbers, and an SDME and LDME must be provided for each term. For hadronic beams $h_1$ and $h_2$, the partonic cross-section is then convolved with the parton density functions (PDF) for the hadrons
\begin{equation}
  \begin{aligned}
    & \dif\sigma(h_1 h_2 \to \snq{\PQ}{2S+1}{L}{J}{} X) = \sum_i \sum _j
    \int \dif[x_1] \int \dif[x_2] \\
    & \qquad f_1(x_1, \pmu, a_i) f_2(x_2, \pmu, a_j) 
    \dif\hat{\sigma}(a_i a_j \to \snq{\PQ}{2S+1}{L}{J}{} X, \pmu) \mmp{,}
  \end{aligned}
  \label{equ:sigma}
\end{equation}
where the first two sums are over all possible partonic combinations $a_i$ and $a_j$. The PDFs for the two hadrons are given by $f_1$ and $f_2$, evaluated at the partonic energy scale $\pmu$ with corresponding momentum fractions $x_1$ and $x_2$.

For production of the \Ppsi at the Large Hadron Collider (LHC), the fixed-order NRQCD partonic cross-section at leading order in \as is given by,
\begin{equation}
  \begin{aligned}
    \dif\hat{\sigma}(\snq{\Pc}{3}{S}{1}{} X) ={}
    & \sno{\snq{\Pc}{3}{S}{1}{}}{3}{S}{1}{1}
    \dif\hat{\sigma}(\Pg\Pg \to \snq{\Pc}{3}{S}{1}{1} \Pg) \\
    & +\, \sno{\snq{\Pc}{3}{S}{1}{}}{1}{S}{0}{8}\bigl(
    \dif\hat{\sigma}(\Pg\Pg \to \snq{\Pc}{1}{S}{0}{8} \Pg)
    + \dif\hat{\sigma}(\Pq\Pg \to \snq{\Pc}{1}{S}{0}{8} \Pq) \\
    & \qquad + \dif\hat{\sigma}(\Pqqbar \to \snq{\Pc}{1}{S}{0}{8} \Pg)\bigr)
    \\
    & +\, \sno{\snq{\Pc}{3}{S}{1}{}}{3}{P}{J}{8}\bigl(
    \dif\hat{\sigma}(\Pg\Pg \to \snq{\Pc}{3}{P}{J}{8} \Pg)
    + \dif\hat{\sigma}(\Pq\Pg \to \snq{\Pc}{3}{P}{J}{8} \Pq) \\
    & \qquad + \dif\hat{\sigma}(\Pqqbar \to \snq{\Pc}{3}{P}{J}{8} \Pg)\bigr)
    \\
    & +\, \sno{\snq{\Pc}{3}{S}{1}{}}{3}{S}{1}{8}\bigl(
    \dif\hat{\sigma}(\Pg\Pg \to \snq{\Pc}{3}{S}{1}{8} \Pg)
    + \dif\hat{\sigma}(\Pq\Pg \to \snq{\Pc}{3}{S}{1}{8} \Pq) \\
    & \qquad + \dif\hat{\sigma}(\Pqqbar \to \snq{\Pc}{3}{S}{1}{8} \Pg)\bigr)
    \mmp{,}
  \end{aligned}
  \label{equ:sigmaHatPsi}
\end{equation}
where the expansion in the relative quark/antiquark velocity is up to $v^2$. Feed-down from \Pchi states contribute significantly to \Ppsi production, with the partonic cross-sections given by
\begin{equation}
  \begin{aligned}
    \dif\hat{\sigma}(\snq{\Pc}{3}{P}{J}{} X) ={}
    & \sno{\snq{\Pc}{3}{P}{J}{}}{3}{P}{J}{1}\bigl(
    \dif\hat{\sigma}(\Pg\Pg \to \snq{\Pc}{3}{P}{J}{1} \Pg)
    + \dif\hat{\sigma}(\Pq\Pg \to \snq{\Pc}{3}{P}{J}{1} \Pq) \\
    & \qquad + \dif\hat{\sigma}(\Pqqbar \to \snq{\Pc}{3}{P}{J}{1} g)\bigr) \\
    & +\, \sno{\snq{\Pc}{3}{P}{J}{}}{3}{S}{1}{8}\bigl(
    \dif\hat{\sigma}(\Pg\Pg \to \snq{\Pc}{3}{S}{1}{8} \Pg)
    + \dif\hat{\sigma}(\Pq\Pg \to \snq{\Pc}{3}{S}{1}{8} \Pq) \\
    & \qquad + \dif\hat{\sigma}(\Pqqbar \to \snq{\Pc}{3}{S}{1}{8} \Pg)\bigr)
    \mmp{,}
  \end{aligned}
  \label{equ:sigmaHatChi}
\end{equation}
at leading order in \as and expanded up to $v$. In \pythia, a complete set of these $2 \to 2$ unpolarised partonic cross-sections is available at fixed order in \as, including for \snq{\PQ}{3}{D}{J}{} states, where \Pbbbar configurations and arbitrary heavy radial excitations are also included, \eg the \Ppsi[2] or \Pups[1], but may need to be configured by the user. All of these partonic cross-sections are $t$-channel exchanges and diverge for low \pt of the quarkonium. These divergences are regularised either with a hard \pt cut-off or, in the case of multiple parton interactions in \pythia, a smooth damping factor.

The LDMEs of the $\mathcal{O}(1)$ leading colour-singlet Fock state, \ie the Fock state with the same quantum numbers as the physical state, can be determined from the wave-function of the physical state at the origin. The LDMEs for the additional Fock states must be determined from fits to data, although velocity scaling rules can be used to relate these LDMEs. Typically, these fits are made to the differential quarkonium production cross-section with respect to the \pt of the quarkonium, as the colour-singlet contributions dominate at low \pt, while the colour-octet contributions dominate at high \pt. Using LDME fits from CDF, the differential cross-section in \pt of the \Ppsi is described well at both the Tevatron~\cite{Cacciari:1995yt, Braaten:1994vv} and the LHC~\cite{Ma:2010jj, Gong:2008ft, Butenschoen:2010rq}. However, these NRQCD predictions are dominated by colour-octet contributions at high \pt, which are typically transversely polarised, unlike colour-singlet contributions which are typically longitudinally polarised. This is in direct conflict with experimental results, which are primarily unpolarised across all \pt of the \Ppsi~\cite{Butenschoen:2012px, CDF:2007msx, D0:2008yos, ALICE:2011gej, CMS:2012bpf, CMS:2013gbz, LHCb:2014brf, LHCb:2013izl}.

The relative isolation of quarkonia can also be measured experimentally, where a jet clustering algorithm such as anti-$k_\text{T}$ can be used to group the quarkonium with surrounding activity, and observables such as $\pt(\text{quarkonium})/\pt(\text{jet})$ can be constructed. For fixed-order NRQCD predictions, quarkonia are expected to be relatively isolated for colour-singlet contributions. Even for fixed-order colour-octet contributions with a parton shower applied similar to that of \pythia, the quarkonia is still expected to be primarily isolated. However, measurements from both LHCb~\cite{LHCb:2017llq} and CMS~\cite{CMS:2019ebt, CMS:2021puf} do not show this isolation, and instead match distributions that are more consistent with production of the quarkonia from fragmentation rather than fixed-order calculations. In \ccite{Bain:2016clc, Bain:2017wvk}, fragmenting jet functions (FJF) calculated via resummation, and similar to the fragmentation probabilities of \cref{equ:sigmaFrag} below, were used to model the LHCb data. However, these calculations are incomplete and do not include all sources of \Ppsi production such as feed-down.

Until this work, \pythia did not include NRQCD-based fragmentation contributions to quarkonium production. This contribution can be written, formally, as,
\begin{equation}
  \begin{aligned}
    & \dif\sigma(h_1 h_2 \to \snq{\PQ}{2S+1}{L}{J}{} X) = \sum_i \sum_j \sum_k
    \int \dif[x_1] \int \dif[x_2] \int \dif[z] \\
    & \qquad f_1(x_1, \pmu, a_i) f_2(x_2, \pmu, a_j) 
    \dif\hat{\sigma}(a_i a_j \to a_k X, \pmu)
    D_{k \to \snq{\PQ}{2S+1}{L}{J}{}}(z, \pmu) \mmp{,}
  \end{aligned}
  \label{equ:sigmaFrag}
\end{equation}
which is analogous to \cref{equ:sigma}, but now uses the fixed-order partonic cross-section \mbox{$\dif\hat{\sigma}(a_i a_j \to a_k X)$} for parton $a_i$ and $a_j$ scattering to produce parton $a_k$ inclusively, multiplied by the the fragmentation probability $D_{k \to \snq{\PQ}{2S+1}{L}{J}{}}(z, \pmu)$, which can be for either the sum of all Fock states or a specific Fock state. The precise definition of $z$ varies within the literature, but generally all definitions reduce to the relative energy fraction carried by the emitted quarkonium in the high-energy limit.

While this formalism can calculate quarkonium production from fragmentation, it cannot produce realistic event topologies that emulate events observed in experiments. Additionally, this type of calculation typically does not include competition between quarkonium states, \ie the fragmentation probabilities do not simply factorize. In this work, we provide an alternative method that overcomes both of these shortcomings, by introducing quarkonium splittings into the \pythia parton shower. This technique is formally equivalent to analytic resummation for inclusive observables, but provides more comprehensive results via a flexible Monte Carlo generator interface. To avoid double counting, this NRQCD parton shower should not be used together with fixed-order quarkonium processes without considering appropriate matching\footnote{Currently fixed-order quarkonium production is automatically switched off in the \pythia multiparton interactions framework when the NRQCD shower is switched on. Future work is required to match this parton shower with fixed-order processes to ensure no double counting.}.

%%%%%%%%%%%%%%%%%%%%%%%%%%%%%%%%%%%%%%%%%%%%%%%%%%%%%%%%%%%%%%%%%%%%%%%%%%%%%%%
\section{Parton Showers}\label{sec:shower}

A complete description of the \pythia parton shower is available in \ccite{Sjostrand:2004ef}. Here, a brief review is given, with a focus on the algorithmic procedure. Parton showers evolve partons produced from fixed-order calculations at high virtuality $\pmu_\max$ to low virtuality $\pmu_\min$ through a series of probabilistic parton emissions. While various parton showers differ in the choice of the evolution variable \pmu, it is illustrative to consider it as an energy-like scale. The probability for parton $a_i$ to not radiate, \ie $a_i \to a_j a_k$, between virtualities $\pmu_1$ and $\pmu_2$ is given by,
\begin{equation}
  \pss[i](\pmu_1, \pmu_2) = \exp \left(-\int_{\pmu_2}^{\pmu_1}
  \dif[\pmu^2] \frac{1}{\pmu^2} \frac{\as(\pmu)}{2\pi}
  \int_{z_\min}^{z_\max} \dif[z] \sum_{j,k} 
  \psk[i \to j k](z, \pmu) \right) \mmp{,}
  \label{equ:sudakov}
\end{equation}
where $z$ is defined as $E_j/E_i$, the fractional energy parton $a_j$ takes from $a_i$, and \psk[i \to j k] is the splitting kernel for the process $a_i \to a_j a_k$. A summation is performed over all available splitting kernels, \eg for a gluon both the splittings $\Pg \to \Pg\Pg$ and $\Pg \to \Pqqbar$ would be summed, where the second splitting is over all kinematically available quark flavours.

The parton shower then proceeds through an iterative algorithm. For a given parton dipole with virtuality $\pmu_1$, a no-emission probability is uniformly sampled, $\pss[i](\pmu_1,\pmu_2) \in \mathcal{U}(0, 1)$, yielding an equation that can be solved for the emission scale $\pmu_2$. In practice, the integrands in \cref{equ:sudakov} are complicated, and so the emission scale is calculated using an overestimate and rejection method called the veto algorithm. In this case, the summed integrands $P_{i \to j k}$ of \cref{equ:sudakov} are replaced by $O_i$,
\begin{equation}
  \pso[i] = \sum_{j,k} \pso[i \to j k] \mmp{,}
  \text{ such that } \pso[i \to j k] \geq \psk[i \to j k](z, \pmu) \mmp{,}
\end{equation}
is valid for all \pmu in the range $\pmu_2$ to $\pmu_1$, and all $z$ in the range $z_\min$ to $z_\max$. If \as is evolved at first order, the new virtuality $\pmu_2$ can be calculated as
\begin{equation}
  \pmu^2_2 = \Lambda^2 \left(\frac{\pmu^2_1}{\Lambda^2}\right)^
  {\pss[i]^{b_0/\pso[i]}} \mmp{,}
  \label{equ:q2Run}
\end{equation}
where $b_0$ is the one-loop $\beta$-function coefficient~\cite{ParticleDataGroup:2020ssz}, $(33 - 2n_f)/6$, and $\Lambda$ is the scale from which \as is being evolved. If, instead, \as is fixed at one value, the new virtuality becomes,
\begin{equation}
  \pmu^2_2 = \pmu^2_1 \pss[i]^{2\pi/(\as\pso[i])} \mmp{.}
  \label{equ:q2Fix}
\end{equation}

After the virtuality $\pmu_2$ is determined according to \cref{equ:q2Run} or \cref{equ:q2Fix}, one of the quarkonium splittings $a_i \to a_j a_k$ is chosen randomly with probability $\pso[i \to jk]/\pso[i]$. A $z$ is then randomly sampled\footnote{This sampling does not need to be uniform, and in many cases is not, as long as the appropriate transformation is applied after sampling.} between $z_\min$ and $z_\max$. The splitting is accepted if,
\begin{equation}
  \frac{\psk[i \to j k](z, \pmu_2)}{\pso[i \to j k]} \geq \mathcal{U}(0, 1)
\end{equation}
where the right-hand side is a uniformly sampled random variable. If the splitting is rejected, then: the algorithm returns to \cref{equ:q2Run} or \cref{equ:q2Fix}; a new $\pmu_1$ is set as the old $\pmu_2$, $\pmu_2 \to \pmu_1$, and a new $\pmu_2$ is selected; and the procedure continues as before. The parton shower continues until $\pmu_2$ drops below the minimum virtuality, $\pmu_\min$. The virtuality variable \pmu in \pythia~\cite{Sjostrand:2004ef} is defined as,
\begin{equation}
  \pmu^2 \equiv \pte[2] = (s - m_r^2)\, z\, (1 - z)
\end{equation}
where $\sqrt{s}$ is the centre-of-mass energy for the dipole, $m_r$ is the mass of the radiating parton in the dipole, and $z$ is a light-cone momentum fraction carried by the emitted parton\footnote{This definition differs from the one used to construct the kinematics of the evolving system.}.

%%%%%%%%%%%%%%%%%%%%%%%%%%%%%%%%%%%%%%%%%%%%%%%%%%%%%%%%%%%%%%%%%%%%%%%%%%%%%%%
\section{Implementation}\label{sec:imp}

A full set of splitting kernels to complement the \Ppsi and \Pchi fixed-order production implemented in \pythia, described in \cref{sec:nrqcd}, were derived from fragmentation functions in the literature similar to those used in \cref{equ:sigmaFrag}. The implemented splitting kernels, as well as their references, are given in \cref{tab:kernels}. The full forms for all the implemented splitting kernels can be found in the references. To illustrate our procedure for including the NRQCD-based fragmentation functions in the \pythia shower, we give the details for the $\PQ \to \snq{\PQ}{3}{S}{1}{1}\PQ$ splitting. The splitting kernel is given as,
\begin{equation}
  \begin{aligned}
    & \psk[\PQ \to \snq{\PQ}{3}{S}{1}{1} \PQ](z, \sqrt{s}) =
    \frac{\as(\pte)16\sno{\snq{\PQ}{2S+1}{L}{J}{}}{3}{S}{1}{1}}{27m_{\PQ}}
    \frac{1}{(s - m_{\PQ}^2)^4} \\
    & \qquad \biggl(s^2 - 2 m_{\PQ}^2 s - 47m_{\PQ}^4
    - (1 - z)(s - m_{\PQ}^2)(s - (m_{\PQ} + m_{\snq{\PQ}{2S+1}{L}{J}{}})^2) \\
    & \qquad
    + 4s(s - m_{\PQ}^2)z(1 - z)/(1 + z)
    + 12(s - m_{\PQ}^2)^2(1 - z)^2z/(1 + z)^2 \\
    & \qquad
    - 4m_{\PQ}^2(s - m_{\PQ}^2)(8 - 7(1 - z) - 5(1 - z)^2)/(1 + z) \biggr) \\
    & \qquad \times (s - m_{\PQ}^2) \mmp{,}
  \end{aligned}
  \label{equ:split3S11}
\end{equation}
where the LDME \sno{\snq{\PQ}{2S+1}{L}{J}{}}{3}{S}{1}{1} depends upon the final physical state \snq{\PQ}{2S+1}{L}{J}{} that this splitting kernel is being used to produce, \eg the \Ppsi as a \snq{\Pc}{3}{S}{1}{} state.

\begin{table}
  \centering
  \caption{Summary of the implemented quarkonium splitting kernels, with their corresponding reference from the literature. The numbers in parentheses indicate the relevant equations from that reference.\label{tab:kernels}}
  \begin{tabular}{ll|ll}
    \multicolumn{1}{c}{splitting}
    & \multicolumn{1}{c|}{reference}
    & \multicolumn{1}{c}{splitting}
    & \multicolumn{1}{c}{reference} \\
    \midrule
    $\PQ \to \snq{\PQ}{1}{S}{0}{1} \PQ$ & \cite{Braaten:1993mp}(19)
    & $\Pg \to \snq{\PQ}{1}{S}{0}{1} \Pg$ & \cite{Braaten:1993rw}(6) \\
    $\PQ \to \snq{\PQ}{3}{S}{1}{1} \PQ$ & \cite{Braaten:1993mp}(14, 15)
    & $\Pg \to \snq{\PQ}{3}{S}{1}{1} \Pg\Pg$ & \cite{Braaten:1995cj}(3, 8) \\
    %$Q \to \snq{\PQ}{1}{P}{1}{1} \PQ$ & \cite{Yuan:1994hn}(16-21) & & \\
    $\PQ \to \snq{\PQ}{3}{P}{J}{1} \PQ$ & \cite{Yuan:1994hn}(16, 23-27, 98, 99)
    & $\Pg \to \snq{\PQ}{3}{P}{J}{1} \Pg$ & \cite{Braaten:1994kd}(8-12) \\
    $\PQ \to \snq{\PQ}{3}{P}{J}{8} \PQ$ & \cite{Yuan:1994hn}(54, 55, 59-61) \\
  \end{tabular}
\end{table}

This expression differs from that given in \ccite{Braaten:1993mp} for a number of reasons. First, the $z$ convention used there is equivalent to $1 - z$ in \pythia, and so a change of variables is made here. An additional factor of $2\pi$ is also included to account for the $1/(2\pi)$ in \cref{equ:sudakov}. There is already a factor of $\as(\pmu)$ in \cref{equ:sudakov}, so only one factor of \as is included, evaluated at the scale \pte; in \ccite{Braaten:1993mp}, both factors of \as are evaluated at a scale of $3m_{\PQ}$. The choice for this scale is discussed in \cref{sec:results}. Finally, the integration variable of \cref{equ:sudakov} is $\pmu^2 \equiv \pte[2]$, rather than $s$ as provided in \ccite{Braaten:1993mp}, so a Jacobian transforming from $s$ to \pte[2] has been introduced:
\begin{equation}
  s = \frac{\pte[2]}{z(1 - z)} + m_{\PQ}^2 \to
  \dif[s] = \frac{1}{z(1 - z)} \dif[\pte[2]] =
  \frac{(s - m_{\PQ}^2)}{\pte[2]} \dif[\pte[2]] \mmp{.}
\end{equation}
Because there is already a factor of $1/\pte[2]$ in \cref{equ:sudakov}, only the additional factor of $(s - m_{\PQ}^2)$ needs to be included, corresponding to the final line of \cref{equ:split3S11}. A similar conditioning must be done for all the splitting kernels of \cref{tab:kernels}.

An overestimate must also be determined for \cref{equ:split3S11}. This can be determined analytically or numerically. For this splitting, we can factorise \cref{equ:split3S11} into a $z$-independent component, and a $z$-dependent component $f(z)$, given by,
\begin{equation}
  \frac{\as(\pte)16\sno{\snq{\PQ}{2S+1}{L}{J}{}}{3}{S}{1}{1}}{27m_{\PQ}}
  \frac{1}{(s - m_{\PQ}^2)} f(z) \mmp{.}
\end{equation}
A factor of $1/(s - m_{\PQ}^2)^2$ is kept in the $z$-dependent component $f(z)$ to ensure it remains dimensionless. The overestimate can then be written as,
\begin{equation}
  \pso[\PQ \to \snq{\PQ}{3}{S}{1}{1} \PQ] = \frac{3}{2}
  \frac{\as\left(\pte[\min]\right)16\sno{\snq{\PQ}{2S+1}{L}{J}{}}{3}{S}{1}{1}}
       {27m_{\PQ}8m_{\PQ}^2} \mmp{,}
\end{equation}
where we have used the numerical knowledge that $f(z) < 3/2$, and we use the parton shower cut-off to evaluate \as. Since $(s - m_{\PQ}^2)$ must always be larger than $(m_{\PQ} + m_{\snq{\PQ}{2S+1}{L}{J}{}})^2 - m_{\PQ}^2$, we replace $(s - m_{\PQ}^2)$ with $8m_{\PQ}^2$ in the numerator of the second term. Again, a similar procedure can be performed for the remaining splitting kernels of \cref{tab:kernels}.

Some of the splitting kernels above deviate significantly enough from the example to warrant special mention. The $\Pg \to \snq{\PQ}{3}{S}{1}{1} \Pg\Pg$ kernel is a $1 \to 3$ splitting, rather than the normal $1 \to 2$, and so the splitting kernel depends upon another kinematic variable, the digluon mass $m_{\Pg\Pg}^{2}$, which must be sampled when generating the final-state kinematics. Another factor of \as enters into this splitting kernel, which is naturally evaluated at a scale of \pte with no choice given to the user. The $\Pg \to \snq{\PQ}{3}{P}{J}{1} \Pg$ splitting kernel diverges rapidly as $\pte[2] \to 0$, and so a \pte[2]-dependent overestimate has been introduced to increase the sampling efficiency for this kernel. Finally, it should be noted that \cref{tab:kernels} does not include any splittings for the \snq{\PQ}{3}{S}{1}{8} Fock state. These must be treated specially as described in the following.

There is one \snq{\PQ}{3}{S}{1}{8} splitting, given by $g \to \snq{\PQ}{3}{S}{1}{8}$. This is unique because the splitting is $1 \to 1$, rather than $1 \to 2$, and arises from a delta function in the splitting kernel. Because the gluon has the same quantum numbers as the \snq{\PQ}{3}{S}{1}{8}, this splitting can be thought of as an off-shell gluon converting directly to a \snq{\PQ}{3}{S}{1}{8} state when the virtuality of the gluon matches the corresponding quarkonium mass. The splitting kernel is then given by,
\begin{equation}
  \begin{aligned}
    \psk[g \to \snq{\PQ}{3}{S}{1}{8}] ={} & \cfrac{
      -\log\left(1 - \as\left(m_{\snq{\PQ}{2S+1}{L}{J}{}}\right)
      \frac{\pi^2(2J + 1)\sno{\snq{\PQ}{2S+1}{L}{J}{}}{3}{S}{1}{1}}
      {12\left(m_{\snq{\PQ}{2S+1}{L}{J}{}}/2\right)^3}\right)}{
      \as\left(m_{\snq{\PQ}{2S+1}{L}{J}{}}\right)\log(1 + \delta m)} \\
    & \times \left\{
    \begin{array}{ll}
      1 & \text{ if } m_{\snq{\PQ}{2S+1}{L}{J}{}}^2 < \pte[2] <
      m_{\snq{\PQ}{2S+1}{L}{J}{}}^2(1 + \delta m) \\
      0 & \text{ otherwise} \\
    \end{array}
    \right. \mmp{,}
  \end{aligned}
  \label{equ:split3S18}
\end{equation}
where $\delta m$ is a small mass splitting, which for this work is set as $0.01~\gev$. The results of the parton shower are not sensitive to the choice of this mass splitting.

%%%%%%%%%%%%%%%%%%%%%%%%%%%%%%%%%%%%%%%%%%%%%%%%%%%%%%%%%%%%%%%%%%%%%%%%%%%%%%%
\section{Results}\label{sec:results}

\begin{figure}
  \centering
  \includegraphics[width=0.49\textwidth]{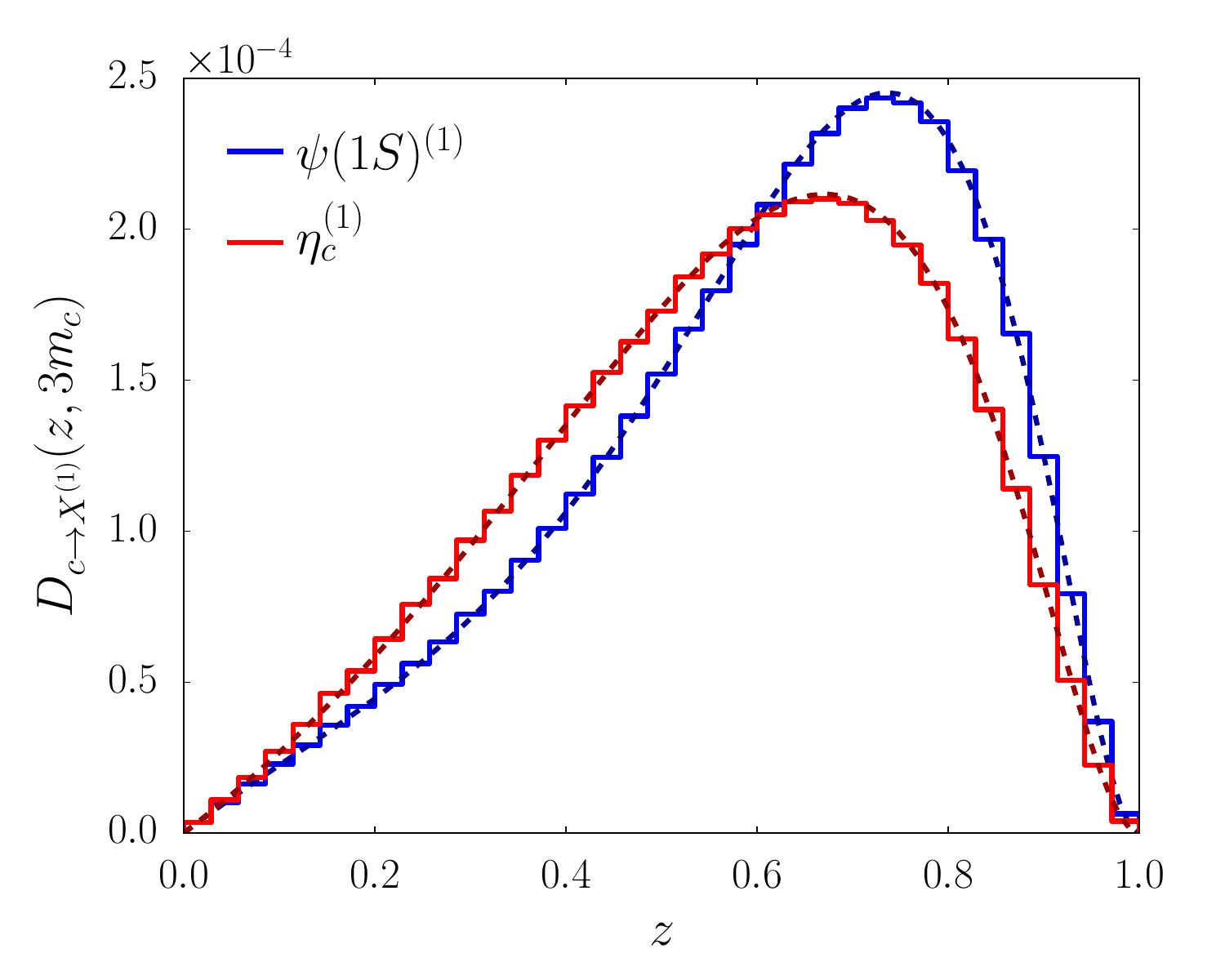}
  \includegraphics[width=0.49\textwidth]{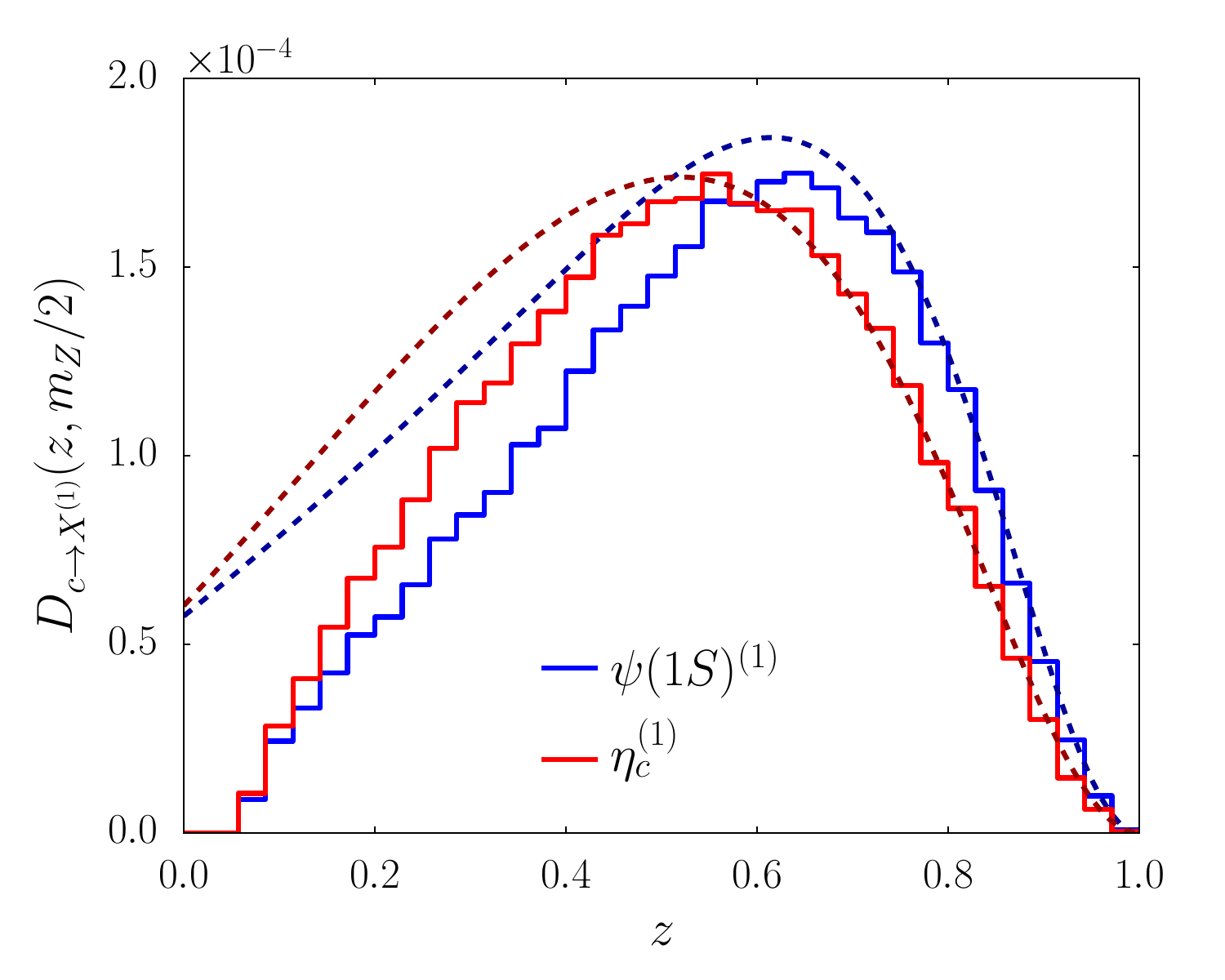}
  \caption{Production of colour-singlet $S$-wave states from charm splittings compared between (solid) \pythia and (dashes) analytic expressions at the energy scales of (left) $3m_{\Pc}$ and (right) $m_\PZ/2$.\label{fig:c1S}}
\end{figure}

The differential fragmentation probability from which the splitting kernel of \cref{equ:split3S11} was derived can be integrated over $s$ to determine the total fragmentation probability. From \ccite{Braaten:1993mp}, this is
\begin{equation}
  \begin{aligned}
    \psd{\Pc \to \Ppsi^{(1)}}(z, 3m_{\Pc}) ={} & \frac{64}{27\pi} \as(3m_c)
    \frac{\sno{\Ppsi}{3}{S}{1}{1}}{m_{\Ppsi}^3} \\
    & \times \frac{z(1 - z)^2(16 - 32z + 72z^2 - 32z^3 + 5z^4)}{(2 - z)^6} \\
  \end{aligned}
  \label{equ:frag3S11}
\end{equation}
for \Ppsi production from the $\Pc \to \snq{\Pc}{3}{S}{1}{1} \Pc$
splitting at the lowest allowed scale of $3m_{\Pc}$. Here, $z$ does not follow the same convention as \cref{sec:shower,sec:imp}, but rather is defined as $2E_{\snq{\PQ}{2S+1}{L}{J}{}}/\com$ where \com is the centre-of-mass energy for the initial dipole. A comparison can be made directly between the parton-shower implementation of this work, and \cref{equ:frag3S11}. This is done by producing a \Pccbar dipole with a sufficiently high \com that mass effects are negligible. The \pythia parton shower, with the inclusion of the individual quarkonium splitting kernels from \cref{sec:imp} with no competition between the quarkonium splittings, is then run on this dipole. All QCD splittings are also switched off. The result can be seen in the left plot of \cref{fig:c1S}, where the \pythia result given by the red histogram matches well with the analytic prediction of \cref{equ:frag3S11} corresponding to the dashed line. Similar agreement is also seen for the \Peta, shown in blue.

\begin{figure}
  \centering
  \includegraphics[width=0.49\textwidth]{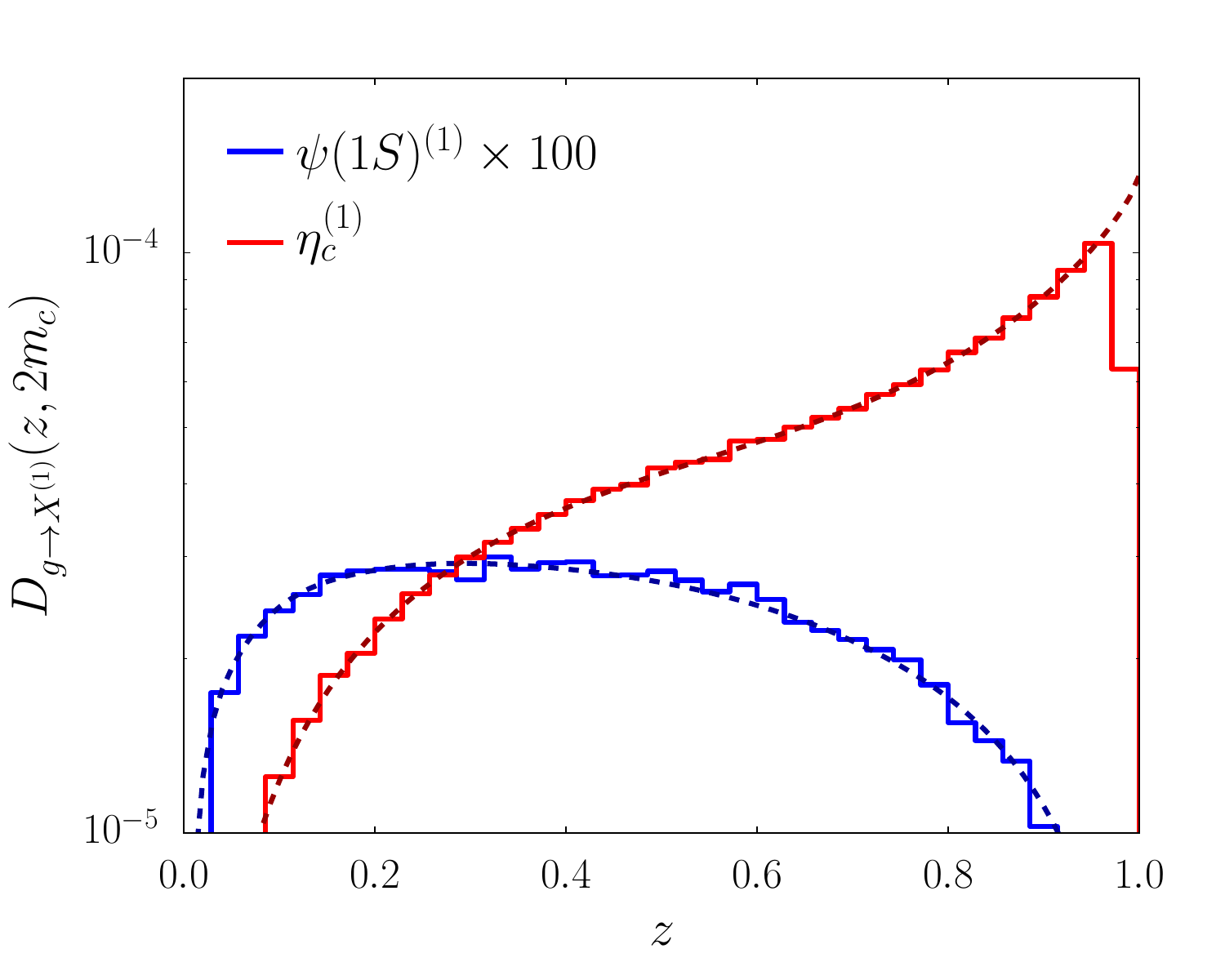}
  \includegraphics[width=0.49\textwidth]{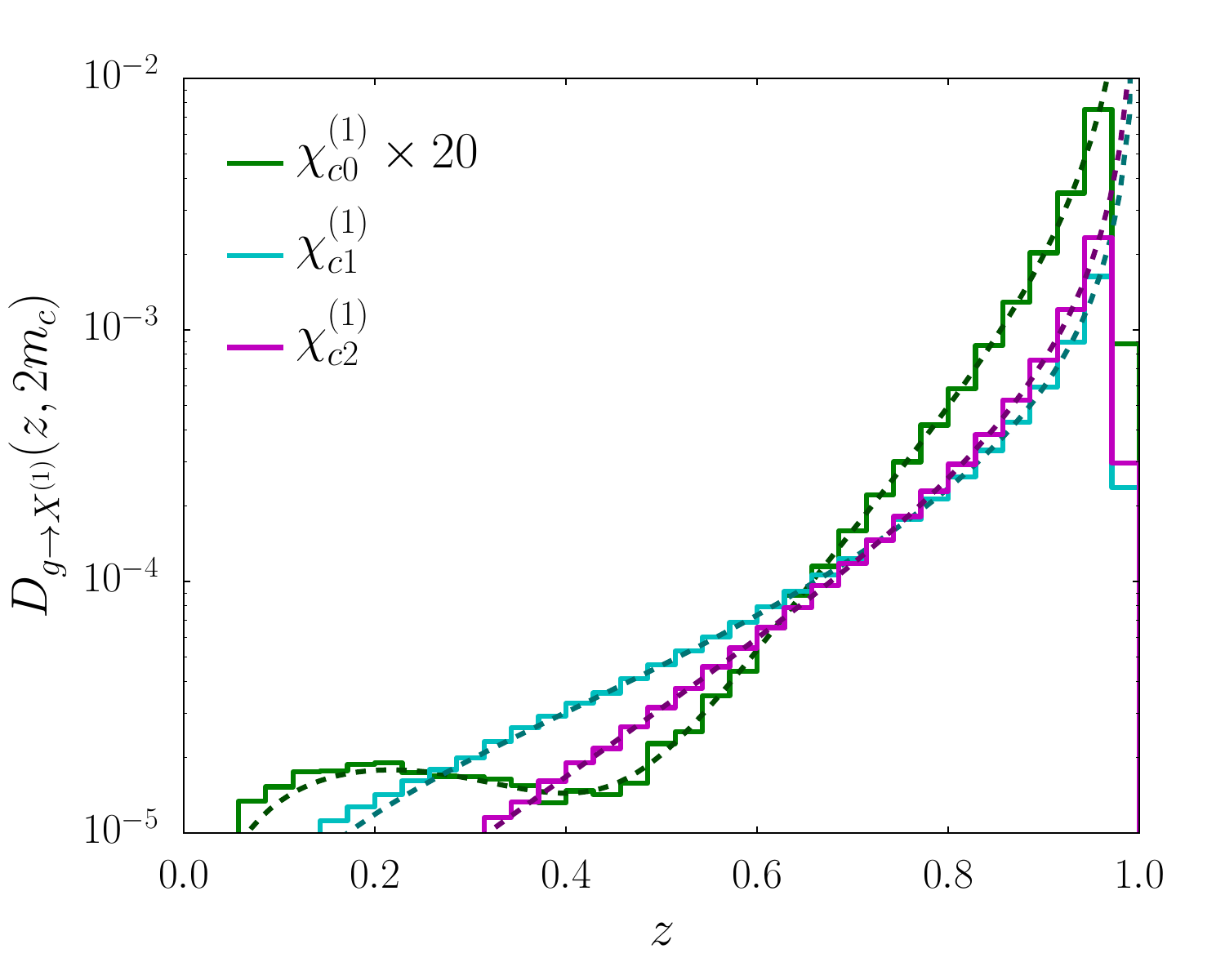}
  \caption{Production of colour-singlet (left) $S$-wave and (right) $P$-wave states from gluon splittings compared between (solid) \pythia and (dashes) analytic expressions at the energy scale of $2m_{\Pc}$.}
  \label{fig:mcg1}
\end{figure}

\begin{figure}
  \centering
  \includegraphics[width=0.49\textwidth]{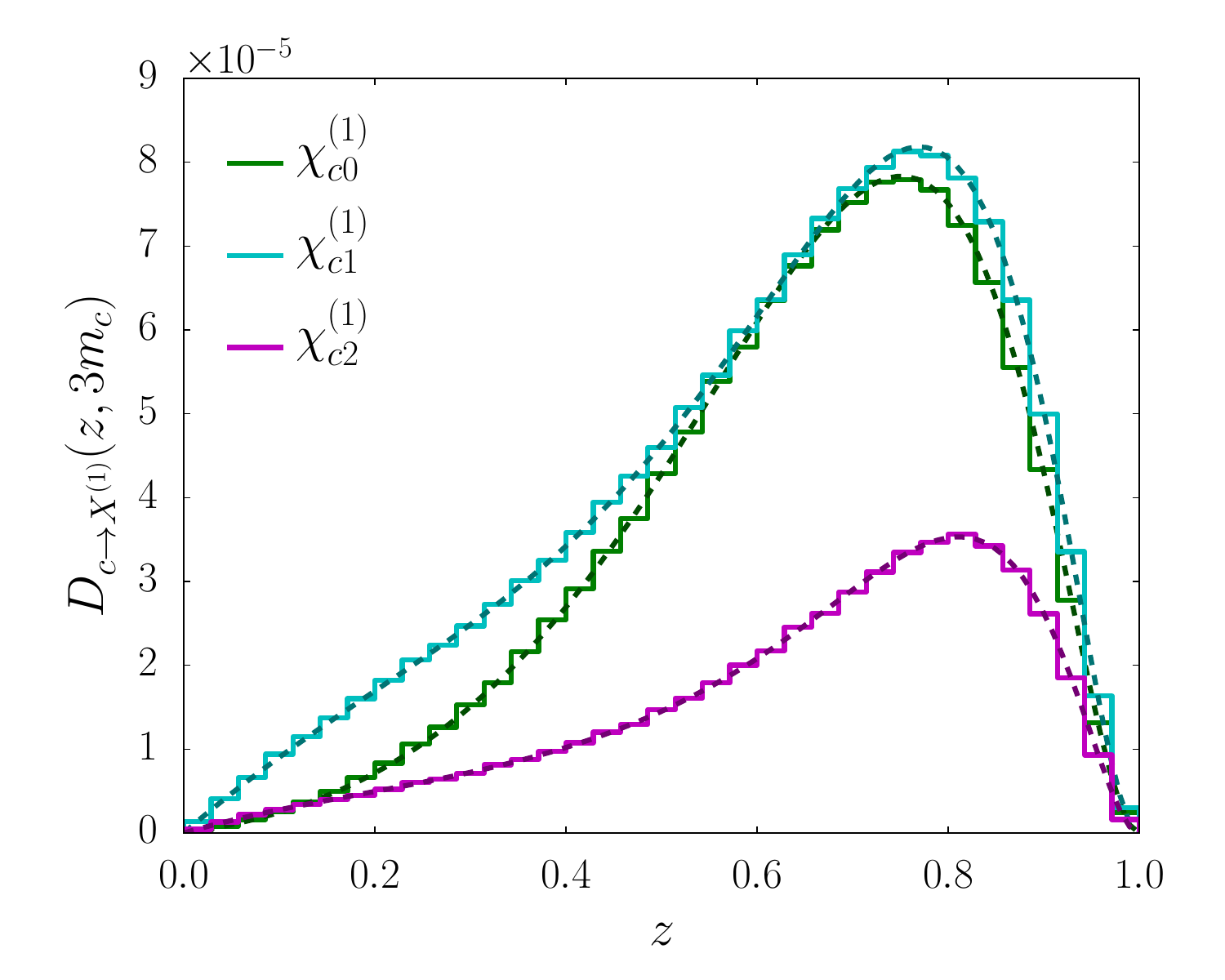}
  \includegraphics[width=0.49\textwidth]{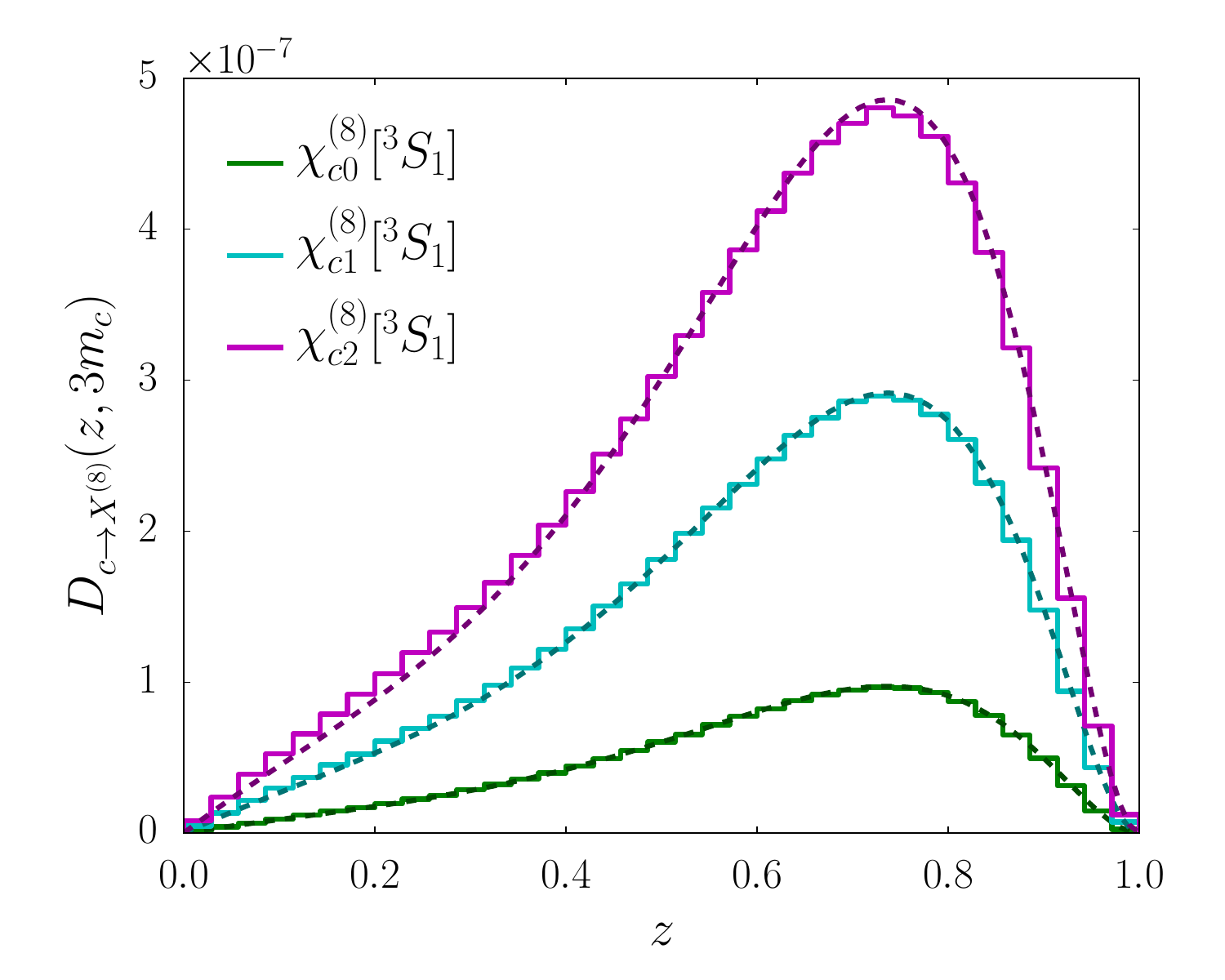}
  \caption{Production of (left) colour-singlet and (right) colour-octet $P$-wave states from charm splittings compared between (solid) \pythia and (dashes) analytic expressions at the energy scale of $3m_{\Pc}$.}
  \label{fig:mcc}
\end{figure}

\begin{figure}
  \centering
  \includegraphics[width=0.49\textwidth]{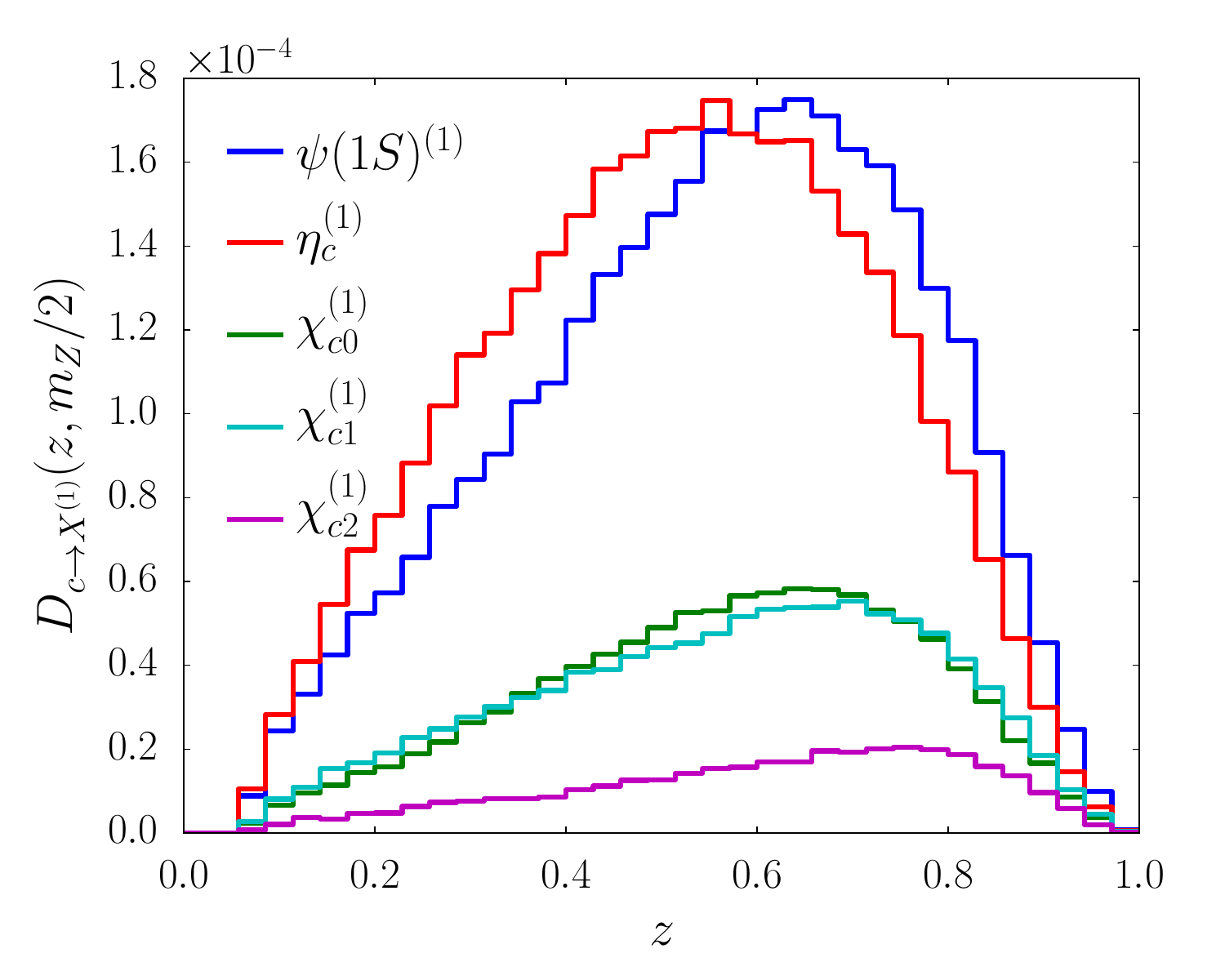}
  \includegraphics[width=0.49\textwidth]{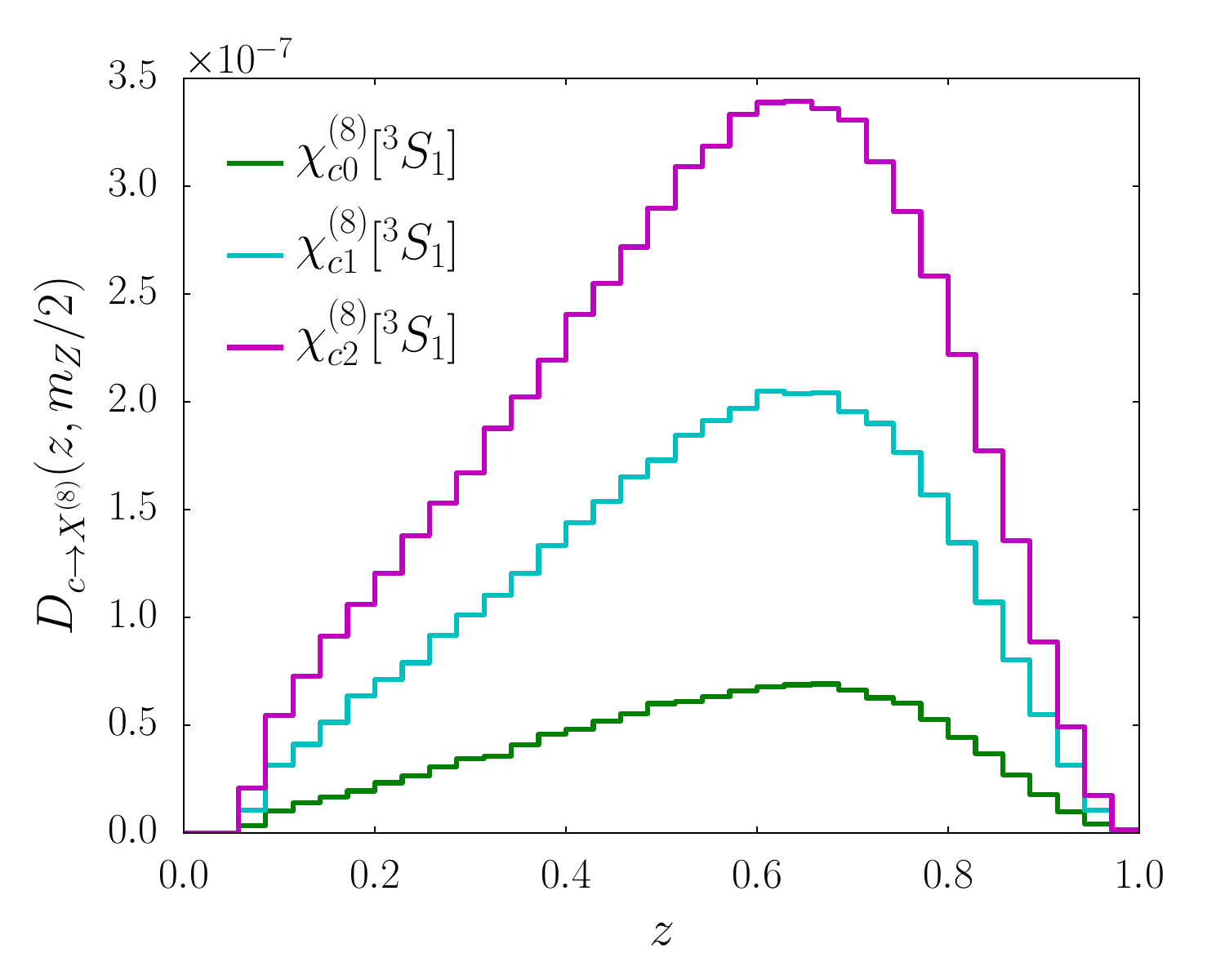}
  \includegraphics[width=0.49\textwidth]{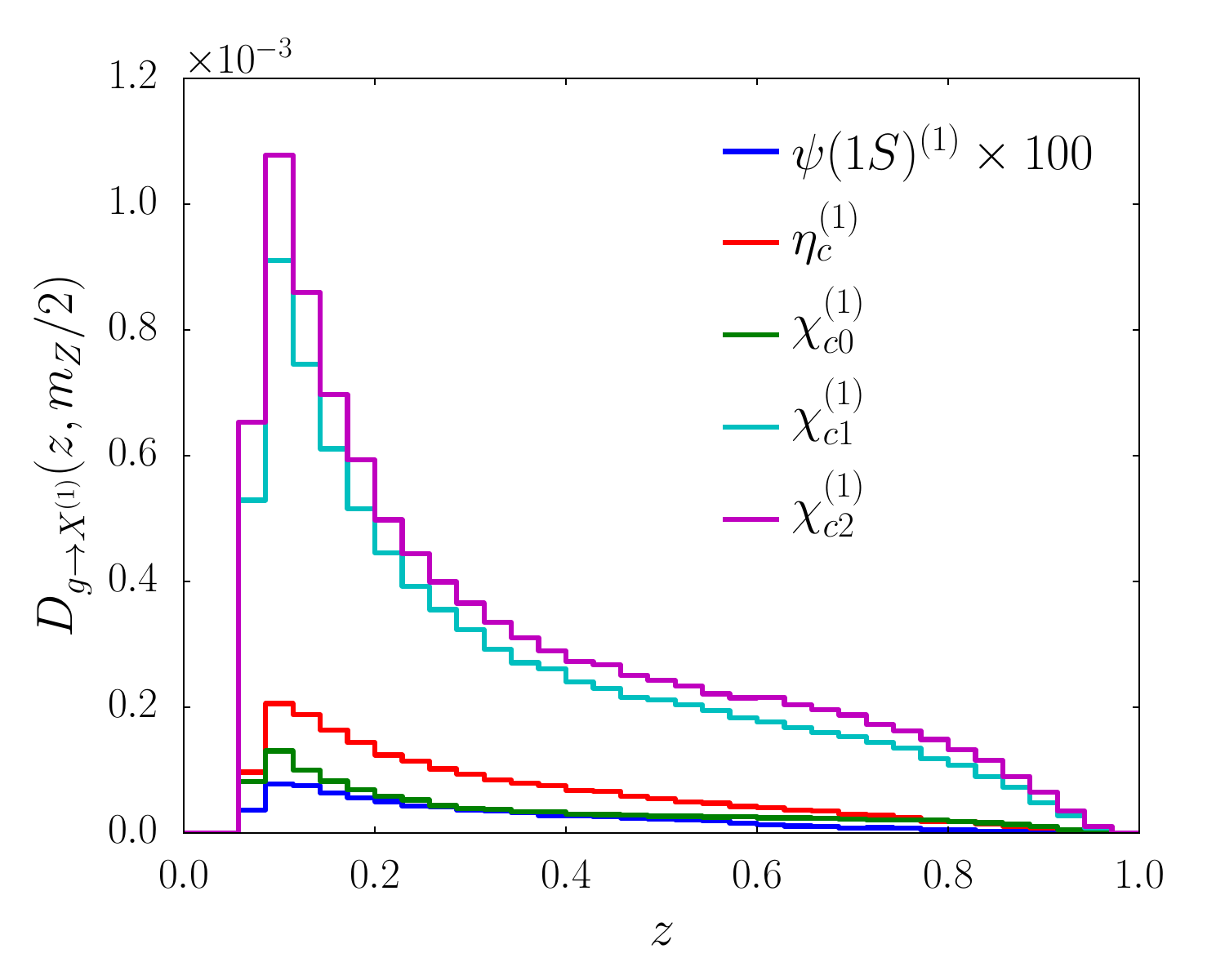}
  \includegraphics[width=0.49\textwidth]{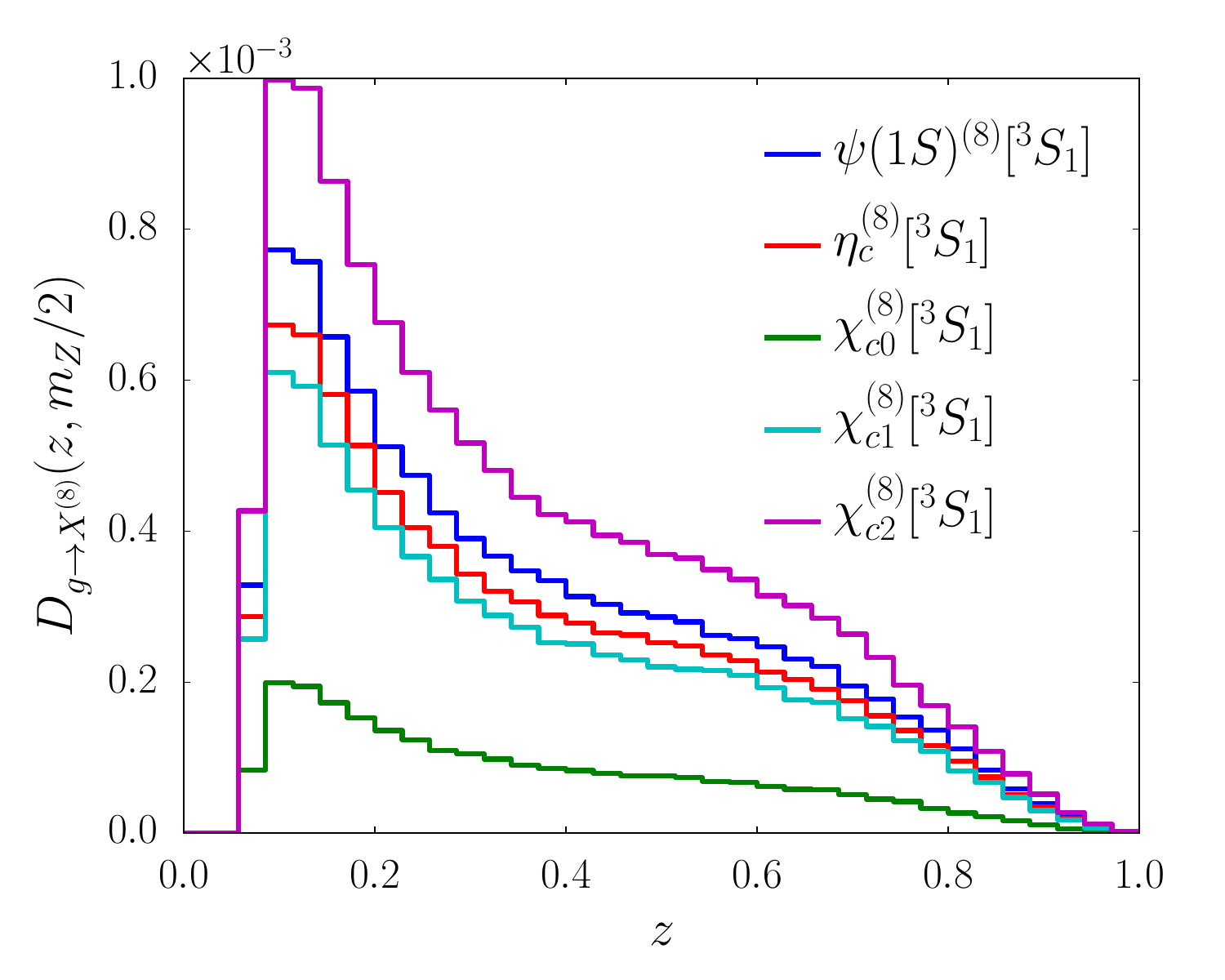}
  \caption{Production of (left) colour-singlet and (right) colour-octet states from (upper) charm and (lower) gluon splittings with \pythia at the energy scale of $m_Z/2$.}
  \label{fig:mz}
\end{figure}

The DGLAP equations can be applied to \cref{equ:frag3S11} to numerically evolve the fragmentation probability from the scale of $3m_{\Pc}$ up to $m_\PZ/2$. In \ccite{Braaten:1993mp} this is provided numerically\footnote{The results are only available in the published version of that paper, not the pre-print, and can be extracted from fig.~2.} for the \Ppsi and \Peta, as given by the dashed red and blue lines, respectively, in the right plot of \cref{fig:c1S}. The implementation of this work can be compared by showering a \Pccbar dipole with $\com = m_\PZ$, and keeping all QCD splittings switched on. The results are also given in the right plot of \cref{fig:c1S}. It should be noted that for high $z$, where mass effects become negligible, the results match well. However, at low $z$ the results differ significantly, because in \ccite{Braaten:1993mp} the masses  of the \Ppsi and \Peta were not considered. Accounting for this mass can be seen clearly in the \pythia result, where there are no quarkonia produced with $z < 2m_{\Ppsi}/m_\PZ$.

Analytic expressions similar to \cref{equ:frag3S11} are available in the literature for all the splittings of \cref{tab:kernels}, evaluated at the lowest kinematically allowed scale. For heavy-quark radiators \PQ, this scale is $3m_{\PQ}$, while for gluon radiators \Pg, the scale is $2m_{\PQ}$. In the left plot of \cref{fig:mcg1} the fragmentation probabilities $\psd{\Pg \to \Ppsi^{(1)}}(z, 2m_{\Pc})$ and $\psd{\Pg \to \Peta^{(1)}}(z, 2m_{\Pc})$ for the colour singlet splittings $\Pg \to \snq{\Pc}{3}{S}{1}{1} \Pg$ and $\Pg \to \snq{\Pc}{1}{S}{0}{1} \Pg$ are given, respectively. There is good agreement between the parton-shower results and the analytic predictions. In the right plot of \cref{fig:mcg1}, comparisons between the parton-shower results and analytic predictions are given for $\psd{\Pg \to \Pchi[\Pc0]^{(1)}}(z, 2m_{\Pc})$, $\psd{\Pg \to \Pchi[\Pc1]^{(1)}}(z, 2m_{\Pc})$, and $\psd{\Pg \to \Pchi[\Pc2]^{(1)}}(z, 2m_{\Pc})$, corresponding to the splittings $\Pg \to \snq{\Pc}{3}{P}{0}{1} \Pg$, $\Pg \to \snq{\Pc}{3}{P}{1}{1} \Pg$, and $\Pg \to \snq{\Pc}{3}{P}{2}{1} \Pg$, respectively. All three parton-shower results agree well with their corresponding analytic predictions.

The fragmentation probabilities, evaluated at the energy scale of $3m_{\Pc}$, for the production of the \Pchi from \Pc-initiated splittings, are given in \cref{fig:mcc}. On the left, the fragmentation probabilities $\psd{\Pc \to \Pchi[\Pc J]^{(1)}}(z, 3m_{\Pc})$ are given for the colour-singlet splittings $\Pc \to \snq{\Pc}{3}{P}{J}{1} \Pc$. On the right, the fragmentation probabilities $\psd{\Pc \to \Pchi[\Pc J]^{(8)}}(z, 3m_{\Pc})$ are given for the colour-octet splittings $\Pc \to \snq{\Pc}{3}{S}{1}{8} \Pc$. All six parton-shower results are in good agreement with the analytic predictions.

Analytic predictions evolved up to the scale of $m_\PZ$ are not available in the literature for any of the processes other than $\Pc \to \snq{\Pc}{3}{S}{1}{1} \Pc$ and $\Pc \to \snq{\Pc}{1}{S}{0}{1} \Pc$ for the \Ppsi and \Peta, respectively. Consequently, we are not able to make comparisons similar to the right plot of \cref{fig:c1S}. We instead provide results at the energy scale of $m_\PZ/2$ in \cref{fig:mz} for all the remaining splittings implemented in \pythia. The fragmentation probabilities for all the colour-singlet $\Pc \to \snq{\Pc}{2S+1}{L}{J}{1} \Pc$ splittings are given in the upper-left plot, while the equivalent fragmentation probabilities for the colour-octet $\Pc \to \snq{\Pc}{3}{S}{1}{8} \Pc$ splittings of the \Pchi states are given in the upper-right plot. The fragmentation probabilities for all the colour-singlet $\Pg \to \snq{\Pc}{2S+1}{L}{J}{1} \Pg$ splittings are given in the lower left, and the fragmentation probabilities for the colour-octet $\Pg \to \snq{\Pc}{3}{S}{1}{8}$ splittings of \cref{equ:split3S18} are given in the lower right.

\begin{figure}
  \centering
  \includegraphics[width=0.49\textwidth]{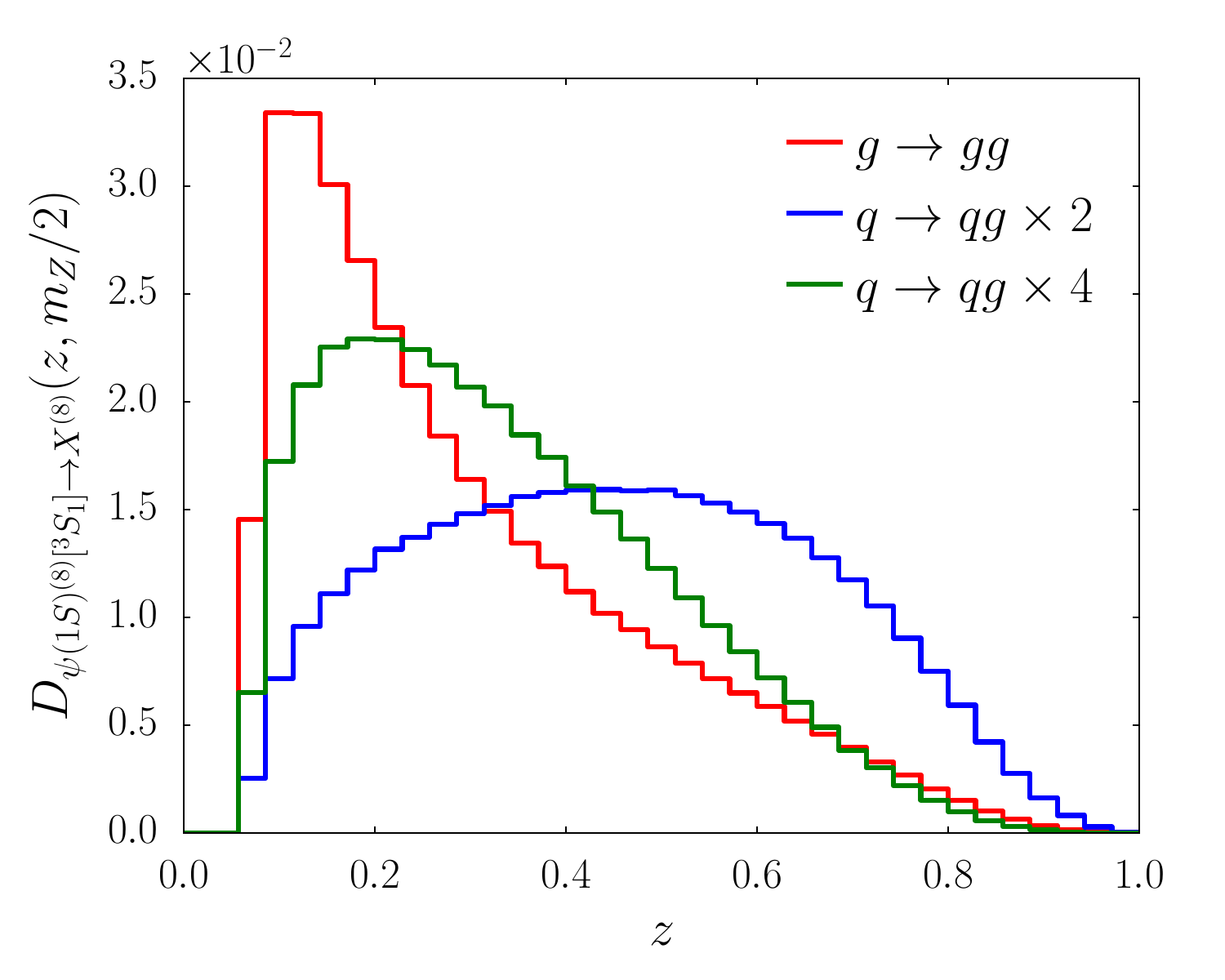}
  \includegraphics[width=0.49\textwidth]{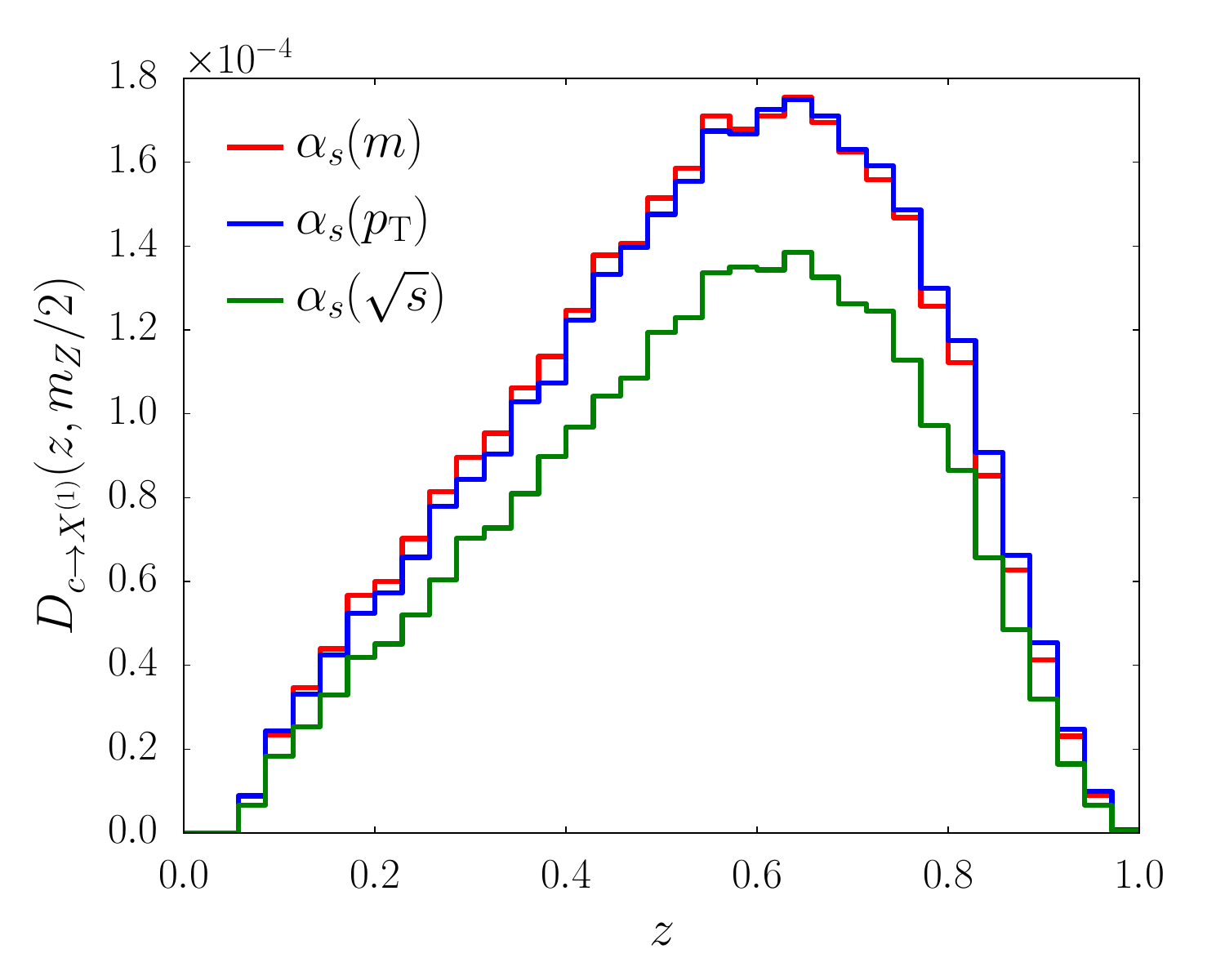}
  \caption{Comparison of (left) splitting kernel choices for the colour-octet \protect\snq{\Pc}{3}{S}{1}{8} splitting with \pythia at the energy scale of $m_Z/2$, and (right) scale choices for the colour-singlet \protect\snq{\Pc}{3}{S}{1}{1} splitting.}
  \label{fig:choice}
\end{figure}

There is some ambiguity in how the colour-octet \snq{\PQ}{3}{S}{1}{8} states, lower-right plot of \cref{fig:mz}, are treated in the parton shower after being produced. Because colour-octet states carry colour charge, they must undergo further QCD radiation before transitioning to their physical state, requiring the emission of one or more soft gluons. The default behaviour in \pythia, prior to this work, was to treat these states with twice the $\Pq \to \Pq \Pg$ splitting kernel,
\begin{equation}
  \psk[\Pq \to \Pq \Pg] \propto \frac{(1 + z^2)}{1 - z} - \frac{2m_{\Pq}^2}{s}
  \mmp{,}
  \label{equ:splitQ}
\end{equation}
\ie the two heavy quarks of the quarkonium state would both be treated as radiators. However, it may be more appropriate to consider this colour-octet state as a massive gluon, where the splitting kernel as derived from \ccite{Kleiss:2020rcg} is
\begin{equation}
  \psk[\Pg \to \Pg \Pg] \propto \frac{2z}{1 - z} -
  \frac{2m_{\Pg}^2}{s} + \frac{4}{3}\bigl((1 - z)/z + z(1 - z)\bigr)
  \mmp{,}
  \label{equ:splitG}
\end{equation}
where the radiating gluon has mass $m_\Pg$ and the emitted gluon is massless. These splittings do not guarantee the gluon emission necessary to transition from an octet to singlet state, so the colour-octet states are required to have slightly larger masses than their corresponding physical states. They are then forced to transition to their physical states through the isotropic emission of a single soft gluon.

The difference between the two splittings of \cref{equ:splitQ,equ:splitG} depends on when the gluon splitting producing the colour-octet state, $\Pg \to \snq{\PQ}{3}{S}{1}{8}$, occurs in the parton-shower evolution. Prior to this splitting, the gluon can continue to radiate, $\Pg \to \Pg \Pg$. Consequently, the second of the two splittings results in a transition much later in the parton-shower evolution, driven primarily by $\Pg \to \Pg \Pg$ splittings. The effects of these two different splitting kernels, as well as the pre-factor used for \cref{equ:splitQ}, can be explored by creating an initial dipole with a $\sn{\Ppsi}{3}{S}{1}{8}$ state at an \com of $m_\PZ$, and evolving this downward to the parton shower cut-off scale. The results for this configuration are shown on the left of \cref{fig:choice}. The difference in the energy radiated away by the colour-octet state depending on the splitting kernel is sizeable. For the previous \pythia default splitting of $\Pq \to \Pq \Pg$ with a pre-factor of $2$, the colour-octet state remains relatively isolated. If the pre-factor is unphysically doubled again to a pre-factor of $4$, the colour-octet state radiates significantly more, as expected. However, for the splitting $\Pg \to \Pg \Pg$, the colour-octet state radiates even more. In this work, this gluon splitting kernel has been used throughout, \ie to produce the bottom-right plot of \cref{fig:mz}.

As mentioned in the discussion of \cref{equ:split3S11}, some choice needs to be made for the scale at which \as is evaluated. Given the diagrams for most quarkonium production, it is natural to choose the first scale as \pte, or whatever the virtuality of the parton shower is. However, the second choice of scale is not as clear. Here, we consider three different options for \pmu: the mass of the quarkonium $m$, the parton-shower virtuality \pte, and the centre-of-mass of the dipole $\sqrt{s}$. In the right plot of \cref{fig:choice}, the impact of this choice is shown for \Ppsi production from a \Pccbar dipole with $\sqrt{s} = m_\PZ$, using only the $\Pc \to \snq{\Pc}{3}{S}{1}{1} \Pc$ splitting kernel. The default choice of \pte is given in blue, and is identical to the corresponding blue curve in the right plot of \cref{fig:c1S}. Switching to $m$ provides a nearly identical result, while using $\sqrt{s}$ significantly reduces the fragmentation probability, although the shape remains similar, as expected. It should be noted for the $\Pg \to \snq{\PQ}{3}{S}{1}{1} \Pg \Pg$ splitting, only one of the three factors of \as is effected by this scale choice, the other two are always evaluated at \pte.

\begin{table}
  \centering
  \caption{Branching fractions for Higgs decays into quarkonia calculated with \pythia. Here, feed-down is from quarkonium decays produced by the parton shower, while hadrons is production from all non-onia hadron decays, \eg from \Pb hadrons. The second set of columns is production from the shower separated by splitting type.\label{tab:brsSplit}}
  \begin{tabular}{r|cc|ccc}
    \toprule
    state & feed-down & hadrons & $X^{(8)}$ & $\Pg \to X^{(1)}$ & $\PQ \to X^{(1)}$ \\
    \midrule
    \Peta & $6.6 \times 10^{-6}$ & $9.5 \times 10^{-3}$ & $2.0 \times 10^{-4}$ & $1.8 \times 10^{-4}$ & $8.0 \times 10^{-5}$ \\
    \Ppsi & $2.0 \times 10^{-4}$ & $1.4 \times 10^{-2}$ & $1.7 \times 10^{-4}$ & $6.5 \times 10^{-6}$ & $7.3 \times 10^{-5}$ \\
    \Pchi[c0] & $6.6 \times 10^{-6}$ & $1.7 \times 10^{-3}$ & $3.0 \times 10^{-5}$ & $2.9 \times 10^{-5}$ & $1.2 \times 10^{-5}$ \\
    \Pchi[c1] & $6.3 \times 10^{-6}$ & $3.2 \times 10^{-3}$ & $8.1 \times 10^{-5}$ & $1.8 \times 10^{-4}$ & $1.0 \times 10^{-5}$ \\
    \Pchi[c2] & $7.1 \times 10^{-6}$ & $1.3 \times 10^{-3}$ & $1.3 \times 10^{-4}$ & $2.1 \times 10^{-4}$ & $3.4 \times 10^{-6}$ \\
    \Peta[b] & $-$ & $-$ & $2.3 \times 10^{-5}$ & $7.4 \times 10^{-6}$ & $4.4 \times 10^{-5}$ \\
    \Pups[1] & $2.4 \times 10^{-5}$ & $-$ & $2.3 \times 10^{-5}$ & $1.7 \times 10^{-7}$ & $4.2 \times 10^{-5}$ \\
    \Pups[2] & $3.8 \times 10^{-6}$ & $-$ & $5.6 \times 10^{-6}$ & $1.0 \times 10^{-7}$ & $2.0 \times 10^{-5}$ \\
    \Pups[3] & $-$ & $-$ & $7.4 \times 10^{-6}$ & $7.0 \times 10^{-8}$ & $1.5 \times 10^{-5}$ \\
    \Pchi[b0] & $1.2 \times 10^{-6}$ & $-$ & $4.8 \times 10^{-6}$ & $3.1 \times 10^{-7}$ & $7.2 \times 10^{-7}$ \\
    \Pchi[b1] & $1.8 \times 10^{-6}$ & $-$ & $1.5 \times 10^{-5}$ & $2.4 \times 10^{-6}$ & $6.7 \times 10^{-7}$ \\
    \Pchi[b2] & $1.9 \times 10^{-6}$ & $-$ & $2.3 \times 10^{-5}$ & $3.2 \times 10^{-6}$ & $3.2 \times 10^{-7}$ \\
    \bottomrule
  \end{tabular}
\end{table}

Up to this point, we have only considered parton-shower configurations where no competition is included between the various quarkonium splittings. This competition is critical when generating comprehensive quarkonium predictions. Because of this competition, the different fragmentation contributions do not simply factorize into independent pieces. To demonstrate the full flexibility of our implementation, we calculate the inclusive branching fractions for standard-model Higgs boson decays into quarkonia. When performing this calculation, all possible quarkonium splitting kernels are switched on with full competition, as well as all other possible splittings, \ie QCD and electroweak. The results are given in \cref{tab:brsSplit}. Here, the branching fractions in the first set of columns is separated into quarkonium production from parton-shower feed-down and from non-quarkonium hadron decays, \eg \Pb hadrons. The second set of columns is production from the parton shower, including feed-down, separated by the splitting type. Here, $X^{(8)}$ is any colour-octet production, $\Pg \to X^{(1)}$ is colour-singlet production from a gluon splitting, and $\PQ \to X^{(1)}$ is colour-singlet production from heavy-quark splitting.

For \Ppsi states, production from non-quarkonium hadron decays dominates the branching fraction, while the parton-shower contribution is nearly two orders of magnitude smaller. This is similar for all the charmonium branching fractions. For bottomonium production there is no production from non-quarkonium hadron decays, as there are not sufficiently high-mass states. For the \Peta[b] and \Pups, the primary production mechanism is colour-singlet production from \Pb splittings, whereas for the \Pchi[b] states, the primary production mechanism is from colour-octet splittings.

%%%%%%%%%%%%%%%%%%%%%%%%%%%%%%%%%%%%%%%%%%%%%%%%%%%%%%%%%%%%%%%%%%%%%%%%%%%%%%%
\section{Conclusions}\label{sec:con}

We have implemented a full-featured quarkonium parton shower in the \pythia Monte Carlo event generator, available from version 8.310 onward. This parton shower can be used to model the LHC quarkonium isolation results, as well as explore the production of quarkonia in general. Our implementation includes all the relevant splitting kernels needed for the production of \snq{\PQ}{1}{S}{0}{}, \snq{\PQ}{3}{S}{1}{}, and \snq{\PQ}{3}{P}{J}{} states with arbitrary radial excitations and for $\PQ \in \Pc, \Pb$. These splitting kernels have been validated against fragmentation probabilities available in the literature, whenever possible, and found to be in good agreement. We provide first predictions for a number of fragmentation probabilities evolved up to the energy scale of $m_\PZ/2$. The framework is flexible, so that an arbitrary number of competing splitting kernels can be included. Various choices are available for the radiation of the colour-octet states, as well as a choice of scale for evaluating \as. Finally, we provide comprehensive first predictions for the production of \snq{\PQ}{1}{S}{0}{}, \snq{\PQ}{3}{S}{1}{}, and \snq{\PQ}{3}{P}{J}{} states from decays of the standard-model Higgs boson. Future work includes developing a method to match these results from the parton shower with fixed-order predictions, including matrix-element corrections, and introducing additional splitting kernels, such as those needed for double-heavy meson and baryon production.

%%%%%%%%%%%%%%%%%%%%%%%%%%%%%%%%%%%%%%%%%%%%%%%%%%%%%%%%%%%%%%%%%%%%%%%%%%%%%%%
\section*{Acknowledgements}
We thank T.~Sj\"ostrand who provided help working with the \pythia parton shower and gave feedback on the manuscript, P.~Skands who provided a thorough code review of our work, C.~Bierlich for performing initial comparisons between the parton-shower and fixed-order calculations in \pythia, and the remainder of the \pythia collaboration for their support of this work. We also thank V.~Bertone, who provided code and explained QCD predictions of the evolution of charm fragmentation to charmonium, S.~Prestel who provided initial impetus to the project, and J.~Zupan and M.~Whitehead who gave feedback on the manuscript. Finally, we thank S.~Rous for providing us with the working name \textsc{Leto} for this project.

SM is supported by the Fermi Research Alliance, LLC under Contract No.~DE-AC02-07CH11359 with the U.S.~Department of Energy, Office of Science, Office of High Energy Physics. PI is supported by NSF grants OAC-2103889 and NSF-PHY-2209769. NC and LL were supported by the MCnetITN3 H2020 Marie Curie Initial Training Network, contract 722104. In addition LL is supported by Swedish Research Council,
contracts numbers 2016-03291 and 2020-04869.

\bibliography{article}

\end{document}